\documentstyle[prd,twocolumn,aps,epsf]{revtex}
\newcommand{\singlefig}[2]{
\begin{center}
\begin{minipage}{#1}
\epsfxsize=#1
\epsffile{#2}
\end{minipage}
\end{center}}
\newcommand{\segmentfig}[3]{
\begin{minipage}{#1}
\epsfxsize=#1
\epsffile{#2}
\begin{center}
{\small \mbox{#3}}
\end{center}
\end{minipage}}
\begin{document}
\draft
\title{The fate of Reissner-Nortstr\"{o}m black hole in 
the Einstein-Yang-Mills-Higgs system}
\author{Takashi Tamaki\thanks{electronic
mail:tamaki@gravity.phys.waseda.ac.jp}
}
\address{Department of Physics, Waseda University,
Shinjuku, Tokyo 169-8555, Japan}
\author{Kei-ichi 
Maeda\thanks{electronic mail:maeda@gravity.phys.waseda.ac.jp} 
}
\address{Department of Physics, Waseda University,
Shinjuku, Tokyo 169-8555, Japan}
\address{Advanced Research Institute for Science and Engineering, 
Waseda University, Shinjuku, Tokyo 169-8555, Japan}
\date{\today}
\maketitle
\begin{abstract}
We study about an evaporating process of black holes 
in SO(3)  Einstein-Yang-Mills-Higgs system. 
We consider a  massless scalar field which couple neither 
with the Yang-Mills field  nor with the Higgs field 
surrounding the black hole. We
discuss differences in evaporating  rate between a monopole
black hole and a Reissner-Nortstr\"{o}m (RN) black hole. 
Since a RN black hole is unstable below the point at which a 
monopole black hole emerges, it will 
transit into a monopole black hole as suggested via a catastrophe theory. 
We then conjecture the following: 
Starting from a Reissner-Nortstr\"{o}m black hole, 
the mass decreases via the Hawking radiation and the black hole 
will reach a critical point. Then it transits to a monopole black hole. 
We find that the evaporation rate will increase continuously  or
discontinuously  according to the type of  phase transition,
that is either  second order or first order, respectively. 
After its transition, the evaporation will never stop 
because the  Hawking temperature of a 
monopole black hole diverges at the zero horizon limit and 
overcomes the decrease of the transmission amplitude $\Gamma$. 
\end{abstract}
\pacs{04.70.-s, 04.40.-b, 95.30.Tg. 97.60.Lf.}
\section{Introduction}                            %
For many years, there have been various efforts to find a  theory
of ``everything".  One of the candidates is the superstring
theory which has been the cause of much attention for last several years in
the context of black hole thermodynamics. Since the discovery of 
black hole radiation by Hawking\cite{Hawking} which is now called
Hawking radiation,  black hole thermodynamics takes its 
position beyond the analogy of ordinary thermodynamics.  

But Hawking radiation is a semiclassical phenomenon which  means 
that space-time itself is treated classically and matter 
field is quantized around its metric.  Although the gravitational
field should also be quantized when the curvature radius gets as small
as  the Plancknian length ($l_{p}\sim 1.6\times 10^{-33}cm$), usually
we ignore it and  estimate the effect of Hawking radiation, e.g.,
as $\gamma$-ray sources of the early universe\cite{Page}.  When
we consider the evaporation process of a Schwarzschild black hole,
the Hawking temperature  arises monotonically and Hawking
radiation does not stop, so classical physics will break down
and  quantum gravity effects should be considered. This is a   serious
unsolved problem which will be a key to quantum gravity. But if
we think about Hawking radiation of a Reissner-Nortstr\"{o}m (RN) black
hole in the Einstein-Maxwell (EM) system, its fate is 
completely different because its temperature will go down and
the evaporation process will cease,  if we assume that the electric
charge is conserved.  

Such fates and related things have been investigated by many 
authors.  But as for the black holes with non-Abelian
hair\cite{Tor2,colored,Torii,DHS,Greene,LM,Tor},  they have not been
much investigated because their black hole solutions 
themselves are only obtained
numerically, which takes much work compared with black holes
analytically obtained.  Another cause is perhaps due to their
instability in which case  the evaporation process need not be considered.  
But for some types of non-Abelian black holes, there
exist  stable solutions. Particularly, a monopole black hole
which is found in  SO(3) Einstein-Yang-Mill-Higgs (EYMH) system
is important in the context of  Hawking
radiation\cite{BFM,Lee,Aichelberg,Tachi}. In the EYMH system, 
if we consider the evaporation process of the RN black hole, its
fate will be rather different from that in the EM system, since it
may experience a phase transition  and become a monopole black
hole. In contrary to the RN black hole, when a monopole  black hole
evaporates, the Hawking temperature rises monotonically like
the Schwarzschild  black hole and it may have the possibility to
become a regular monopole. If this is the case,  it may shed new
light on the problem of the remnant of Hawking radiation.  So we need
to study its evaporation.  

This paper is organized as follows.  In Sec. II, we introduce 
basic Ans\"{a}tze and the field  equations in the EYMH system. 
In Sec. III, we briefly review black holes in the EYMH system and 
their thermodynamical properties. In 
Sec. IV, we investigate the evaporating features of RN and
monopole  black holes in the EYMH system. In Sec. V, we make some 
concluding remarks and mention some discussion. 
Throughout this paper we use  units $c=\hbar =1$. Notations and definitions 
such as Christoffel symbols  and curvature follow
Misner-Thorne-Wheeler\cite{MTW}.   
\section{Basic equations}    %
We consider black hole spacetime in the SO(3) EYMH model as
\begin{eqnarray}
S=\int d^{4}x\sqrt{-g}\left[\frac{R}{2\kappa^{2}} + L_{m}\right],
\label{EYMH}
\end{eqnarray}
where $\kappa^{2} \equiv 8\pi G$ with $G$ being Newton's
gravitational constant. $L_{m}$ is  the Lagrangian density of 
matter fields which  are written as 
\begin{equation}
L_{m}  =  -\frac{1}{4}F^{a}_{\mu\nu}F^{a\mu\nu}
-\frac{1}{2}(  D_{\mu}\Phi^{a}  )(  D^{\mu}\Phi^{a}  )
-\frac{\lambda}{4}(\Phi^{a}\Phi^{a}-v^{2})^{2}\ .
\label{matter lag}     
\end{equation}
$F^{a}_{\mu\nu}$ is the field strength of SU(2) YM field 
and is expressed by its potential $A^{a}_{\mu}$ as
\begin{equation}
F^{a}_{\mu\nu} =  \partial_{\mu}A^{a}_{\mu}-
\partial_{\nu}A^{a}_{\nu}+e \epsilon^{abc}A^{b}_{\mu}A^{c}_{\nu},
\label{B6}     
\end{equation}
with a gauge coupling constant $e$.
$\Phi^{a}$ is a real triplet Higgs field and $D_{\mu}$ is the covariant 
derivative:
\begin{equation}
D_{\mu}\Phi^{a} =  \partial_{\mu} \Phi^{a}+
e \epsilon^{abc}A^{b}_{\mu}\Phi^{c}.
\label{B7}
\end{equation}
The theoretical parameters $v$ and $\lambda$ are a 
vacuum expectation value  and a self-coupling 
constant of a Higgs field, respectively.  To obtain black hole 
solutions,  we assume that a space-time is static and spherically
symmetric, in which case  the metric is written as 
\begin{equation}
ds^{2}=-f(r)e^{-2\delta (r)}dt^{2}+
f(r)^{-1}dr^{2}+r^{2}d\Omega^{2},
\label{B8}
\end{equation}
where $f(r) \equiv  1-2Gm(r)/r$. 
For the matter fields, we adopt the hedgehog ansatz given by 
\begin{equation}
\Phi^{a}  =  v \mbox{\boldmath $r$}^{a} h(r), 
\label{B12} 
\end{equation}
\begin{equation}
A^{a}_{0}=0, \label{B13}
\end{equation}
\begin{equation}
A^{a}_{\mu}  =  \omega^{c}_{\mu}\epsilon^{acb} 
\mbox{\boldmath $r$}^{b}
\frac{1-w(r)}{er}, \;\;\;\;\; (\mu=1,2,3),
\end{equation}
where $ \mbox{\boldmath $r$}^{a} $ and $\omega^{c}_{\mu}$ 
are a unit radial vector in the internal space and a triad, 
respectively. 

Variation of the action  (\ref{EYMH}) with the matter Lagrangian 
(\ref{matter lag}) leads to the field equations
\begin{equation}
\frac{d\delta}{d\tilde{r}} = -  8\pi\tilde{r}\tilde{v}^{2}
\tilde{K}  ,
\label{MPBD4-1} 
\end{equation}
\begin{equation}
\frac{d\tilde{m}}{d\tilde{r}}= 
  4\pi\tilde{r}^{2}\tilde{v}^{2}  
\left[
f\tilde{K}+\tilde{U}
\right],
\label{MPBD4-2}
\end{equation}
\begin{equation}
\frac{d^{2}w}{d\tilde{r}^{2}} =  
\frac{1}{f}
\left[
\frac{1}{2}\frac{  \partial \tilde{U}  }{  \partial w  }
+  8\pi\tilde{r} \tilde{v}^{2}\tilde{U} 
\frac{dw}{d\tilde{r}} 
-\frac{  2\tilde{m}  }{  \tilde{r}^{2}  }\frac{dw}{d\tilde{r}}   
\right],
\label{MPBD4-3}
\end{equation}
\begin{equation}
\frac{d^{2}h}{d\tilde{r}^{2}} = -\frac{dh}{d\tilde{r}}  
\frac{1}{\tilde{r}}  
+\frac{1}{f}\left[   \frac{  \partial \tilde{U}  }{  \partial h  }
+  8\pi\tilde{r} \tilde{v}^{2}\tilde{U}\frac{dh}{d\tilde{r}}
-\frac{1}{\tilde{r}}\frac{dh}{d\tilde{r}}  
\right],
\label{MPBD4-4}  
\end{equation}
where
\begin{equation}
\tilde{U} \equiv  
\frac{(1-w^{2})^{2}}{2\tilde{r}^{4}}
+\left(  \frac{ wh }{ \tilde{r} }  \right)^{2}
+\frac{  \tilde{\lambda}  }{4}(h^{2}-1)^{2},
         \label{MPpotential}
\end{equation}
\begin{equation}
\tilde{K} \equiv    
\frac{1}{\tilde{r}^{2}}\left(  \frac{dw}{d\tilde{r}}  \right)^{2}
+\frac{1}{2}
\left(  \frac{dh}{d\tilde{r}}  \right)^{2}  .     
\label{MPkinetic}     
\end{equation}
We have introduced the following dimensionless variables:
\begin{equation}
\tilde{r}=evr, \ \  \tilde{m}=Gevm, \ \  
\label{B14} 
\end{equation}
and dimensionless parameters:
\begin{equation}
\tilde{v}=v \sqrt{G}, \;\; \tilde{\lambda}=\lambda/e^{2}.
\label{B15}
\end{equation}
Although the solution exists when $v\leq M_{Pl}$, where $M_{Pl}$ 
is Planck mass, it can be described 
by a classical field configuration in the limit of a weak gauge 
coupling constant $e$, because its Compton wavelength $\sim e/v$
is much smaller than the radius of the  classical monopole
solution ($\sim 1/ev$) in this case. Moreover, since the energy density is 
$\sim e^{2}v^{4} \ll M_{Pl}^{4}$, we can treat this classically 
if we ignore the effect of  gravity.  The boundary conditions at
spatial infinity are
\begin{equation}
 m (\infty )= M< \infty,\ \ \delta (\infty )=0,\ \ h(\infty )=1,
\ \ w(\infty )=0.    \label{B16}
\end{equation}
These conditions imply that space-time approaches a flat 
Minkowski space with  a charged object.

To obtain a black hole solution, we assume
the existence of a regular event horizon at $r=r_H$. So the 
metric components are 
\begin{equation}
m_{H}\equiv m(r_H)=\frac{r_{H}}{2G},\ \   
\delta_{H}\equiv\delta(r_H)<\infty.
\end{equation}
We also require that no singularity exists outside the
horizon, i.e., 
\begin{eqnarray}
m(r)<\frac{r}{2G}\  ~~~~~{\rm  for} ~~~ r>r_{H}\ . \label{B11}
\end{eqnarray}
For the matter fields, the square brackets in 
Eqs.~(\ref{MPBD4-3})-(\ref{MPBD4-4}) 
must vanish at the horizon. Hence we find that 
\begin{equation}
\left. \frac{dw}{d\tilde{r}} \right|_{ \tilde{r}=\tilde{r}_{H} }
=\frac{  w_{H}  }{F}
\left(1-w^{2}_{H}-h^{2}_{H}\tilde{r}^{2}_{H}  \right)       
\label{MPBD4-6}   
\end{equation}
\begin{equation}
\left.\frac{dh}{d\tilde{r}}\right|_{  \tilde{r}=\tilde{r}_{H}  }
=-\frac{  h_{H}  }{F}
\left[2w^{2}_{H}+\tilde{\lambda}\tilde{r}^{2}_{H}(h^{2}_{H}-1 ) 
\right] 
\label{MPBD4-7}    
\end{equation}
where
\begin{eqnarray}
F&=& 2\pi\tilde{v}^{2}\tilde{r}_{H}\left[2\tilde{r}^{-2}_{H}(
    1-w^{2}_{H}    )^{2}+4w^{2}_{H}h^{2}_{H}
+\tilde{\lambda}\tilde{r}^{2}_{H} (    h^{2}_{H}-1  
)^{2}\right]    \nonumber  \\
&&-\tilde{r}_{H}  
\label{bunbo}  .
\end{eqnarray}
Hence, we should determine the values of $w_{H}$,  and $h_{H}$ 
iteratively so that the boundary conditions at infinity are 
fulfilled. 

Non-trivial solution does not necessarily exist for given 
physical parameters. However, for arbitrary values of 
$\tilde{v}$ and $\tilde{\lambda}$, there exists an RN black  hole
solution such as 
\begin{eqnarray}
w\equiv 0 ,\ h\equiv 1 ,\ \delta \equiv 0,\ 
\tilde{m}(\tilde{r})\equiv \tilde{M}-
\frac{2\pi\tilde{v}^{2}}{\tilde{r}}\ .
\label{RN1} 
\end{eqnarray}
$\tilde{M}$ is the gravitational mass at spatial infinity  and
$\tilde{Q} \equiv 2\sqrt{\pi}\tilde{v}$ is the magnetic charge of 
the black hole. The radius of the event horizon of the
RN black hole is constrained to be $\tilde{r}_H \geq \tilde{Q}$.
The equality implies an extreme solution.

Around these black holes, we consider a neutral and massless 
scalar field which does not  couple with the matter fields, i.e.,
either Yang-Mills nor Higgs fields. This is described by the
Klein-Gordon  equation as 
\begin{eqnarray}
\Phi_{,\mu}^{\ ;\mu}=0 .
\label{Klein-Gordon} 
\end{eqnarray}
The energy emission rate of Hawking radiation is given by 
\begin{eqnarray}
\frac{dM}{dt}=-\frac{1}{2\pi}\sum_{l=0}^{\infty}(2l+1)
\int_{0}^{\infty}\frac{  \Gamma(\omega)\omega  }{  
e^{\omega/T_{H}}-1  }d\omega ,
\label{emission} 
\end{eqnarray}
where $l$ and $\Gamma(\omega)$ are the angular momentum and 
the transmission probability in a scattering problem for the 
scalar  field $\Phi$. $\omega$ and $T_{H}$ are the energy of the
particle and the Hawking temperature  respectively. We define as
$\Xi \equiv -dM/dt$.  

The Klein-Gordon equation (\ref{Klein-Gordon}) can be made 
separable,  and we should only solve the radial equation 
\begin{eqnarray}
\frac{ d^{2}\chi }{ d\tilde{r}^{\ast 2} }+\chi\left[ 
\tilde{\omega}^{2} -\tilde{V}^{2} \right]=0 , 
\label{radial} 
\end{eqnarray}
where 
\begin{eqnarray}
\tilde{V}^{2}&\equiv &\frac{f}{\tilde{r}e^{2\delta}}
\left\{ \frac{l(l+1)}{ \tilde{r} }-f\delta^{\prime}+
\frac{2(\tilde{m}-\tilde{m}^{\prime}\tilde{r})}{\tilde{r}^{2}}
 \right\}  , 
\label{potential}  \\
\frac{d\tilde{r}}{d\tilde{r}^{\ast}}&\equiv & f e^{-\delta} ,
\label{tortus}
\end{eqnarray}
where $^{\prime}$ denotes $d/d\tilde{r}$ and $\chi$ is only the 
function of $r$.   We need the normalization as
$\tilde{\omega}=\omega /ev$, $\tilde{V}=V/ev$.  The transmission
probability $\Gamma$ can be calculated by  solving the radial
equation numerically under the boundary condition 
\begin{eqnarray}
\chi \rightarrow Ae^{-i\omega r^{\ast}}+Be^{i\omega r^{\ast}}
\ \ (r^{\ast} \rightarrow \infty ) ,
\label{infty}  \\
\chi \rightarrow e^{-i\omega r^{\ast}}
\ \ (r^{\ast} \rightarrow -\infty ) ,
\label{horizon} 
\end{eqnarray}
where $\Gamma$ is given as  $1/|A|^{2}$. In our case, if we 
obtained the black hole solution,  i.e., the shooting parameters
$w_{H}$ and $h_{H}$, we should integrate (\ref{radial})  and
(\ref{MPBD4-1})-(\ref{MPBD4-4}) simultaneously. 

\section{Black Holes in the Einstein-Yang-Mills-Higgs system} 
In this section, we briefly explain about black holes in 
the EYMH system, particularly about thermodynamical properties. 
We show the relation between horizon radius 
$\tilde{r}_{H}$ and gravitational mass $\tilde{M}$ in Fig. 1(a). 
We denote RN and monopole black holes by a dotted line and 
a solid line, respectively. We chose as $\tilde{\lambda}=\tilde{v}=0.1$. 
In this figure, we can see that RN and monopole black holes emerge 
at the point $B$ which does not change even if we change 
$\tilde{\lambda}$. On the contrary, if we change $\tilde{v}$, 
the point $B$ moves 
and disappears for large $\tilde{v}$ (i,e., they do not emerge.). 
The precise are shown in \cite{BFM}. We only consider the parameter 
region where the point $B$ exists, because it is shown in \cite{Lee} 
that the RN black hole becomes unstable via 
linear perturbation below the point $B$ and later analysis showed that 
the RN black hole may transits to the monopole black hole only in this 
case.

Before denoting the stability of monopole black hole, we show 
the magnification of Fig. 1 (a) around the point $B$ in Fig. 1 (b). 
We find a cusp 
structure at the point $A$ which exists for $\tilde{\lambda}<
\tilde{\lambda}_{crit}$. $\tilde{\lambda}_{crit}$ slowly depends on 
$\tilde{v}$. We can understand them via swallow tail 
catastrophe\cite{Tachi}. We can see in Fig. 1 (b) that for some mass range 
which corresponds to $B$ to $A$,  
there appear three types of solutions
(stable RN black hole, stable and  unstable monopole black holes) which
suggests the violation of weak no hair conjecture. By contrast, 
for $\lambda>\lambda_{crit}$, a cusp structure never 
appears and  the monopole black hole solution merges with the RN black hole
at the point $B$.  From the analysis in \cite{Tachi}, we can summerize 
the stability of black holes as follows: 

(i) As for the RN black hole, the stability changes at the point $B$. 
It is stable or unstable according to be above or below the point $B$. 

(ii) As for the monopole black hole, when $\tilde{\lambda}<
\tilde{\lambda}_{crit}$, it is unstable along the curve from $A$ to $B$, 
otherwise it is stable. When $\tilde{\lambda}>\tilde{\lambda}_{crit}$, 
it is always stable. 

We also show the inverse temperature $1/T_{H}$ in terms of the 
gravitational mass $M/(M_{pl}^{2}/ev)$  in Fig. 1(c). Assuming
the conservation  of charge and starting with the RN black hole,
the point $B$ is a key to the fate of the  black hole. Because if
we consider the RN black hole in the EM system, the RN black hole is 
always stable and  the evaporation will cease at the extreme
limit because the temperature vanishes there.  While if we have the 
RN black hole in the EYMH system, the RN black hole becomes
unstable below the  critical point $B$ and will change to a
monopole black hole by either second- or first-  order
transition  according to $\lambda>\lambda_{crit}$ or $\lambda
<\lambda_{crit}$.  After this transition, because the temperature
diverges to infinity at the $r_{H} \rightarrow 0$  limit, so we may
not stop evaporating a monopole black hole and find that one of 
the candidates for the remnant is a self-gravitating monopole. 
We show the diagram for $\tilde{\lambda}=\tilde{v}=0.1$. 
But the results are qualitatively same for other parameters 
(see Fig. 9 (a) in \cite{Tachi}). 

The criterion stable or unstable is based on 
casastrophe which coincide with the analysis via linear perturbation. 
So if we think evaporation process and time evolution of them, 
the results may change. But it may be laborious to calculate 
such evolution, here we consider mainly about 
transmission amplitude of 
a scalar field assuming black holes as a background. 

\section{Evaporation of Black Holes in the Einstein-Yang-Mills-Higgs system} 

Even if we assume that the background spacetime is fixed and 
ignore backreaction, it is not evident to predict the final fate of 
black holes. Naively speaking, the temperature is the main cause 
to decide evaporation process. But the transmission amplitude 
$\Gamma$ may also affect the results. For example, a dilatonic 
black hole in the EM-dilaton system has different properties  via
a coupling constant 
$\alpha$ of the dilaton field to the matter field\cite{GM}. 
If $\alpha >1$, the temperature of the black hole diverges and 
the effective  potential $V^{2}$ grows infinitely high
simultaneously at the extreme limit\cite{footnote1}.  In this
case, it is not evident how to decide  whether or not the emission
rate diverges. In \cite{JK}, it turned out that the divergence of
the temperature at the extreme limit overcomes that of the
effective potential, resulting in a divergence of the  emission
rate.  

In the case of a monopole black hole, it is not even evident whether or 
not the effective potential  diverges at the $r_{H}\rightarrow 0$ 
limit, because its solution is only obtained numerically. 
Fig. 2 shows the effective potential $(V/ev)^{2}$ in terms of 
the radial coordinate for $v/M_{pl}=0.05$, $\lambda /e^{2}=1$ and 
$r_{H}/(ev)^{-1}=0.1$, $0.3$. Taking the horizon radius as smaller, 
the potential $V^{2}$ becomes  larger and our numerical results
suggest that its potential diverges within that limit, so we must
analyze the emission rate to decide  whether or not the
evaporation will stop.  Another interesting point is how
Hawking radiation changes at the point $B$ in the transition
process  of an RN black hole to a monopole black hole. 
Near the point $B$ we can suggest something
concrete under some assumptions.  For this, we assume the
following: (i) The coupling constant $e$ is small enough so
that we can treat the gravitational field classically at the
point B. (ii) A discharge process does not occur during the
evaporation\cite{Zaumen}.  (iii) The coupling of the matter fields
(YM and Higgs fields) to the scalar field even if it exists,  the
results would not be much affected. The last ansatz may seem to be
strong, but it might turn out to be true near the point $B$
because the monopole  black hole around there is very close to the RN
black hole, which is a vacuum solution. In fact the field
strength of YM field for such a monopole black hole is much
smaller than that in the  other cases. 

Before seeing such properties, we show the transmission 
probability $\Gamma$  of a RN black hole and a monopole black
hole in terms of $\omega$ for the horizon radius
$r_{H}/(ev)^{-1}=0.55$ and $v/M_{pl}=0.05$, $\lambda /e^{2}=0.1$,
$1$  in Fig. 3 (a).  We depict only $l=0$, $1$ modes. The RN black
hole,  monopole black hole with $\lambda /e^{2}=0.1$, monopole
black hole with $\lambda /e^{2}=1$   correspond to the dotted
lines, the solid lines and the dot-dashed lines, respectively. 
As we see the dominant contribution for the Hawking radiation is
$l=0$,  because the contributions of the higher modes are  suppressed
by the centrifugal barrier. In what follows, we will ignore
the contributions from $l\geq 2$.  Although $\Gamma$ is the
largest for an RN black hole, these are almost
indistinguishable.   So, when evaluating the value of the
emission rate of a monopole black hole,  we may conclude that
$\Gamma$ may not be the main origin of the difference
from that of the RN black hole.  However, for a monopole black hole
with a smaller horizon radius, its difference from an RN black hole
becomes clear.  Fig. 3 (b) shows the same diagram in Fig. 3 (a)
with the same parameters $v$ and $\lambda$  but with a smaller
horizon radius $r_{H}=0.3/ev$. We can see the difference clearly. 
It is because
the size of non-trivial structure   becomes larger compared with
the horizon radius for the monopole black hole of smaller
horizon. 
We also show the examples of the radiation spectrum $-d^{2}M/dtd\omega$ 
in terms of $\omega$ in Fig. 4 (a), (b). The parameters correspond 
to Fig. 3 (a), (b). One can see the reason why we can ignore the 
contribution for $l\geq 2$. Actually, the difference caused by 
it is below $1\%$ from one of our calculation. 

We return to the first concern, i.e., what happens when the horizon 
radius changes via Hawking radiation.  In Fig. 5 (a),  we show the
emission rate $\Xi$ in terms of the gravitational mass $M$ of an 
RN black hole  and a monopole black hole for $v/M_{pl}=0.05$,
$\lambda /e^{2}=1$.  The difference between an RN black hole and 
a monopole black hole may be caused by the Hawking temperature
$T_{H}$, because the emission rate $\propto T_{H}^{4}$. 
In Fig. 5 (b), we show the emission rate $\Xi$ in terms of 
the horizon radius $\tilde{r}_{H}$ for the same parameters 
in Fig. 5 (a). This diagram strongly suggest that the evaporation  
will not stop even at the
$r_{H}\to 0$ limit.  This resembles the situation
of the dilatonic black hole at the extremal limit, i.e., whether
or not evaporation stop depends only on $T_{H}$\cite{JK}.   
Thus whenever we think black hole space-times as background, 
the transmission amplitude does not change the scenario estimating from 
the temperature. 

We
estimate the time scale of the evaporation using $T_{time} \equiv
M/\Xi$ as the indicator,  and show this time scale in terms of
the gravitational mass $M$ in Fig. 6 for the same solutions  in
Fig. 5 in the CGS units.  Near the bifurcation point $B$,
$T_{time} \sim 10^{-37}/e^{3} \ second$, and below the
bifurcation point, the time evolution of the monopole black hole
is completely  different from that of an RN black hole.   

Next, we study the evaporation rate near the bifurcation point 
when either the first or second order  transition to the monopole
black hole occurs. In Fig. 7, we show the emission rate $\Xi$ in 
terms of the gravitational mass $M$ of RN and monopole  black holes 
for $v=0.05 M_{pl}$, $\lambda /e^{2}=0.1$,
$0.2$, $0.3$, $0.4$, $1$ near  the bifurcation point. The curves
from $B$ to $A$ correspond to the emission rate of the unstable
branch.  In the case that the transition is first order, the
emission rate will also change discontinuously as shown by an arrow in
Fig. 7. 

It may be interesting to ask in which direction the RN black hole 
jumps to the monopole black hole  or how much black hole entropy
(or radius of event horizon) will increase after the phase
transition.  In order to analyze this problem properly, we have
to include a back reaction effect of  Hawking radiation, which is
very difficult and has not yet been solved.  However, we may find
some constraints through the following considerations.  Because the RN
black hole emits particles, it will lose some of the
gravitational mass.  But, whether the horizon radius increases
or not may depend on two time scales, i.e. the
evaporation time and the transition time. If we apply the
catastrophe theory,  the entropy of the black hole will increase
and  the horizon radius will increase.  But this
analysis is based on the assumption that the change of  black
hole states can be treated adiabatically.   Since  the coupling
constant $e$ is so small that  each state can be described by a
quasi-stationary solution, we may expect that  the horizon
radius will increase after the transition. 

We also confirm that  an other choice of values of
$v$  and $\lambda$ does not provide any serious difference 
in the evaporation process. In Fig. 8, we show  one of
the interesting cases ($\lambda /e^{2}=0.1$ and $v=0.2 M_{pl}$) 
where the bifurcation point $B$ is very near the extreme RN
solution.  In this case, this diagram suggests that the RN black
hole first almost ceases the evaporation process  and becomes close
to the extreme one, and then it transits to a monopole black hole
and will start to  evaporate again. In other parameters, 
these diagrams are basically similar to the above cases: 

(i) The emission rate diverges at the $\tilde{r}_{H}\to 0$ limit. 

(ii) Near the bifurcation point $B$, the emission rate changes 
continuously or discontinuously according to $\tilde{\lambda}$ is 
above or below $\tilde{\lambda}_{crit}$. 

\section{Conclusion and discussion}    %
 
We have considered an evaporation process of 
the RN and monopole black holes in the EYMH system. We have analyzed 
a real massless scalar field  which couple to neither the NA
field nor the Higgs field.  We may suggest how RN  and monopole
black holes evolve through an evaporation process in the EYMH system.  
We have the following results. 

(i) We investigated the evaporation process of an RN black hole, 
in particular, near the bifurcation point  where this merges with
a branch of monopole black holes. Since the RN black hole becomes
unstable  there, we expect that it transits into a monopole
black hole. This transition will be first- or second-order
according to  whether $\lambda $ is smaller than some critical
value
$\lambda_{crit}$ or not. We show that  the evaporation rate
changes continuously or discontinuously depending on whether the
transition which occurs is second- or first-order. Our results suggest
that the Hawking radiation near the bifurcation point is
determined  only by the temperature of the black hole.  
We can understand this as follows. 
Because in particular, around this region, we find little difference in 
the transmission probability  between a monopole black hole and 
an RN black hole. 

(ii) When the horizon radius becomes small, the transmission probability 
of a monopole black hole  becomes small compared with that of an 
RN black hole.  Though it can not stop evaporating in our
analysis because the increase of temperature of a monopole black
hole  at the $r_{H}\rightarrow 0$ limit,  quantum effects of
gravity may cause a serious effect  on it and would   overcomes
the decrease of $\Gamma$. 
 
We finally remark on some subjects which we leave to the future.
When  we consider the fate of an RN black hole via Hawking
radiation, we may take into account the effects of charge loss if
it is to be expected,  and have to include a coupling to the YM field
or Higgs field before we  consider the effects of quantum
gravity.  The second is the concern with the critical behavior\cite{Chop}.
There are several works about it in the  EYM or Einstein-Skyrme
systems\cite{Chop-YM,Skyrme}, in which the Schwarzschild black hole is the
most stable one.  But in the EYMH system, since a monopole black
hole becomes more stable than the RN  black hole below a certain 
critical mass, it would be interesting to study the critical
behavior in  the EYMH system.  
Finally, it may be more interesting to look for the ``real"
critical behaviour in our present phase transition via
Hawking evaporation.
Those are under investigation. 

\section*{ACKOWLEDGEMENTS}
Special thanks to J. Koga, T. Torii and T. Tachizawa for useful 
discussions.  T. T is thankful for  financial support from the
JSPS. This work was supported partially by a 
JSPS Grant-in-Aid (No. 106613), 
and by the Waseda University Grant  for Special
Research Projects.


\begin{figure}[htbp]
\segmentfig{8cm}{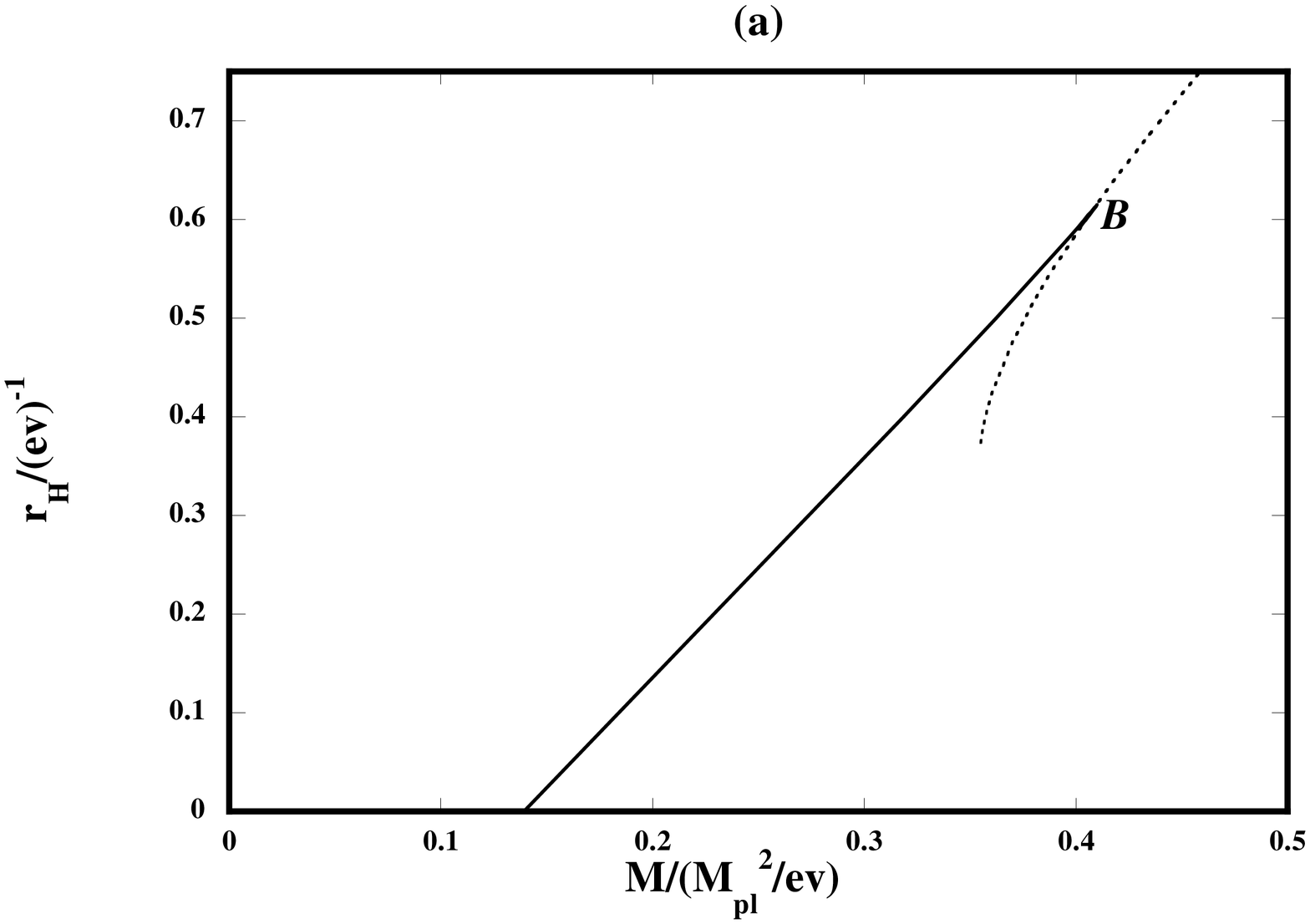}{}  \\
\segmentfig{8cm}{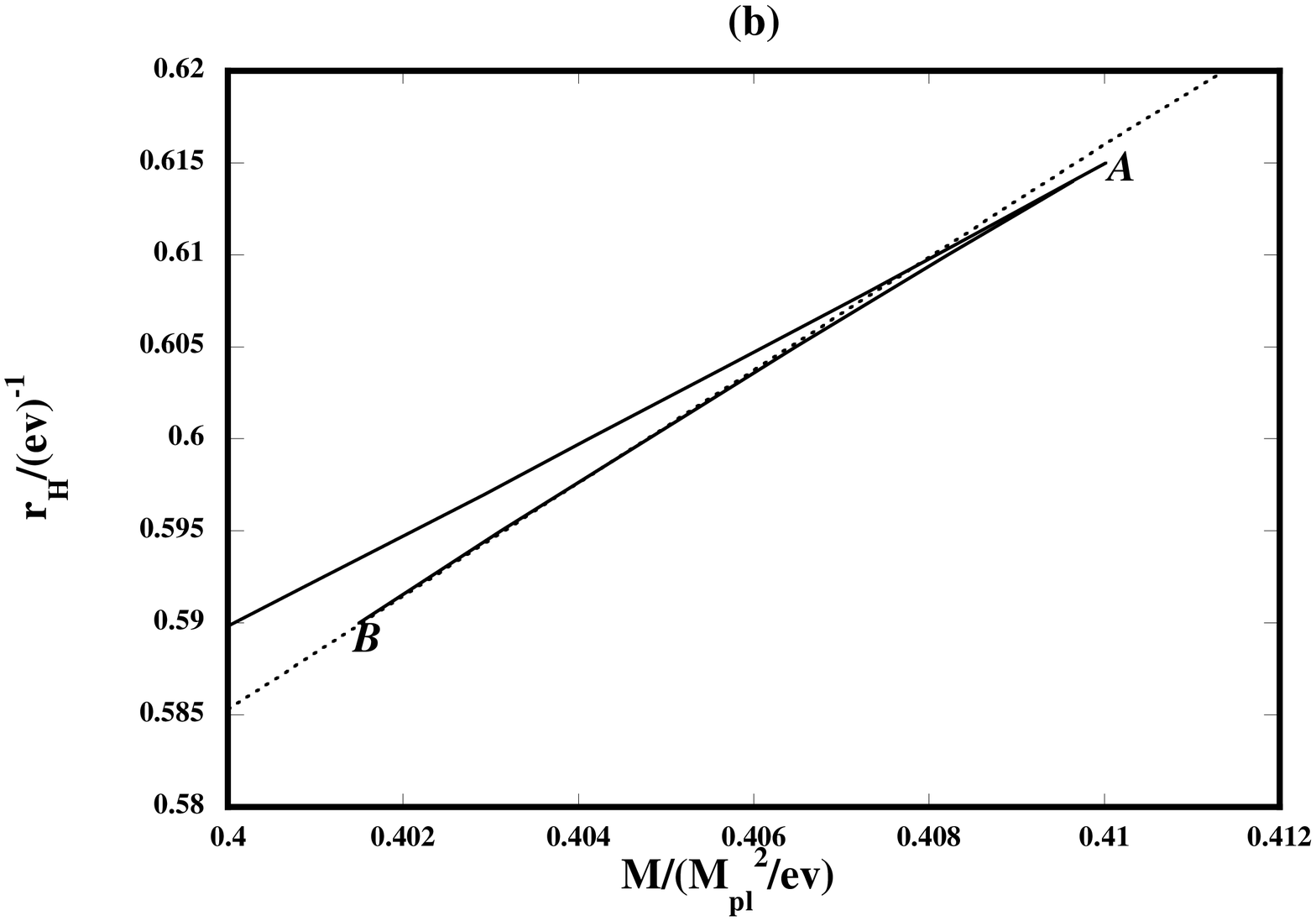}{}  \\
\segmentfig{8cm}{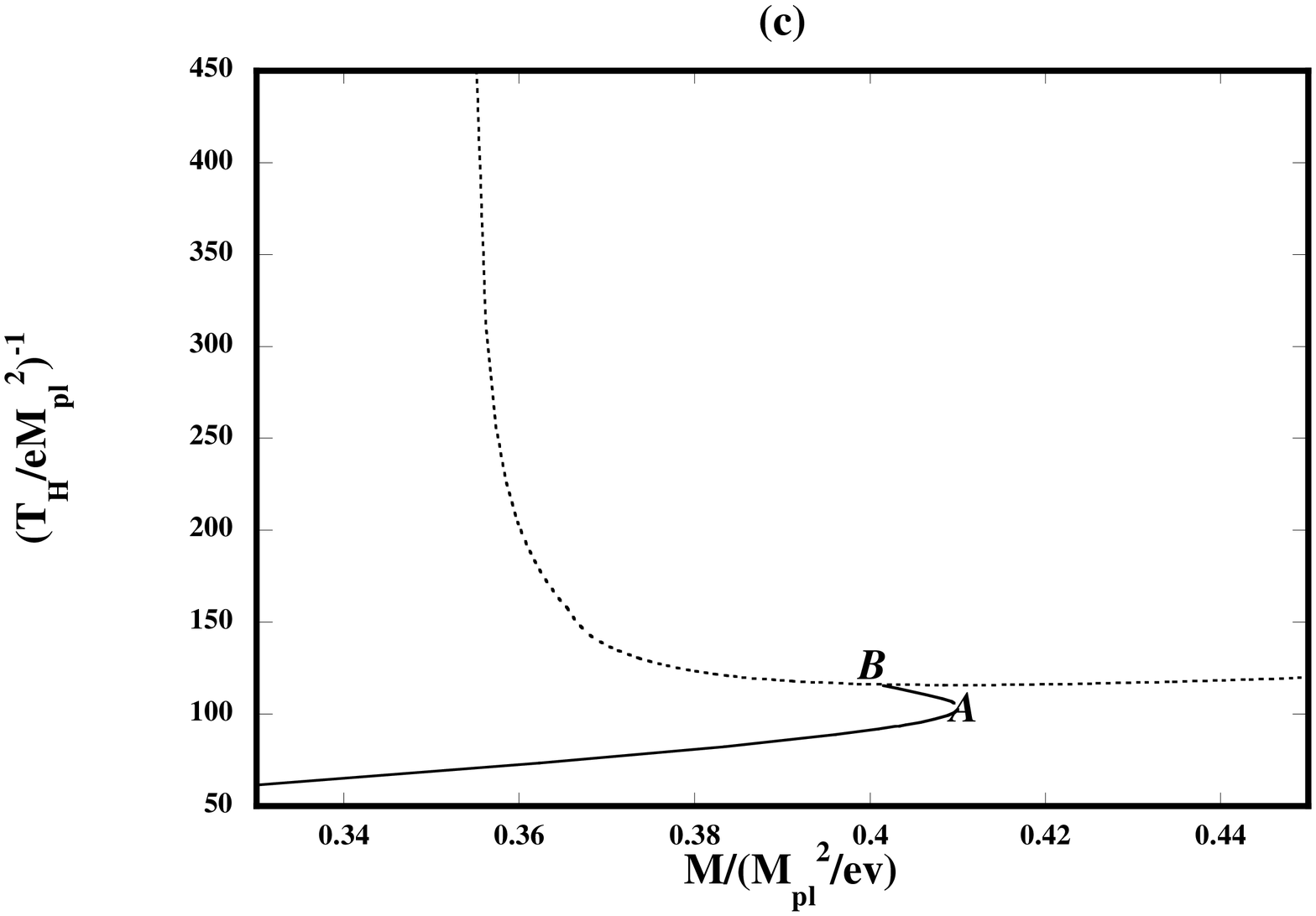}{}
\caption{(a) the gravitational mass $M/(M_{pl}^{2}/ev)$ and the horizon 
radius $r_{H}/(ev)^{-1}$ relation and (b) its magnification diagram near 
the point $B$ and (b) the inverse Hawking
temperature $(T_{H}/eM_{pl}^{2})^{-1}$ in terms of the
gravitational  mass $M/(M_{pl}^{2}/ev)$ of the monopole  black
hole with $\lambda /e^{2}=0.1$ (solid lines) and of the RN black hole (dotted
lines). We choose $v/M_{pl}=0.1$ in these diagrams. At the point
$B$,  the RN black hole becomes unstable and change to monopole
black hole. This process is first or second  order correspond to
$\lambda$ is larger than $\lambda_{crit}$ or not. Fig. (c) shows that 
temperature of the monopole black hole diverges at the $\tilde{r}_{H}
\to 0$ limit like the Schwarzschild one. Though we showed the one example, 
which is irrelevant to the results (see Fig. 9 (a) in \protect\cite{Tachi}.). 
\label{monopole} 
}
\end{figure}
\begin{figure}
\begin{center}
\singlefig{8cm}{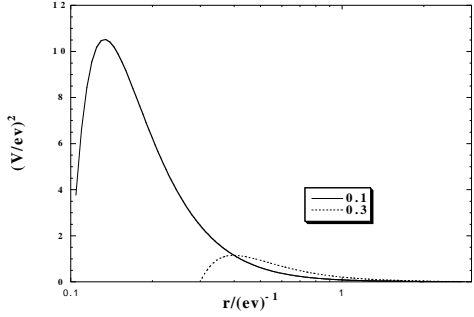}
\caption{We show the potential $V^{2}/(ev)^{2}$ in terms of the radial 
coordinate with 
$\lambda /e^{2}=1$, $v/M_{pl}=0.05$ and $r_{H}/(ev)^{-1}=0.1$, $0.3$ for 
$l=0$ mode. It shows that when horizon radius $r_{H}$ 
becomes small, potential barrier becomes large. 
It seems that if we take $r_{H}\rightarrow 0$, potential barrier diverges.  
\label{potential-fig}
}
\end{center}
\end{figure}
\begin{figure}[htbp]
\segmentfig{8cm}{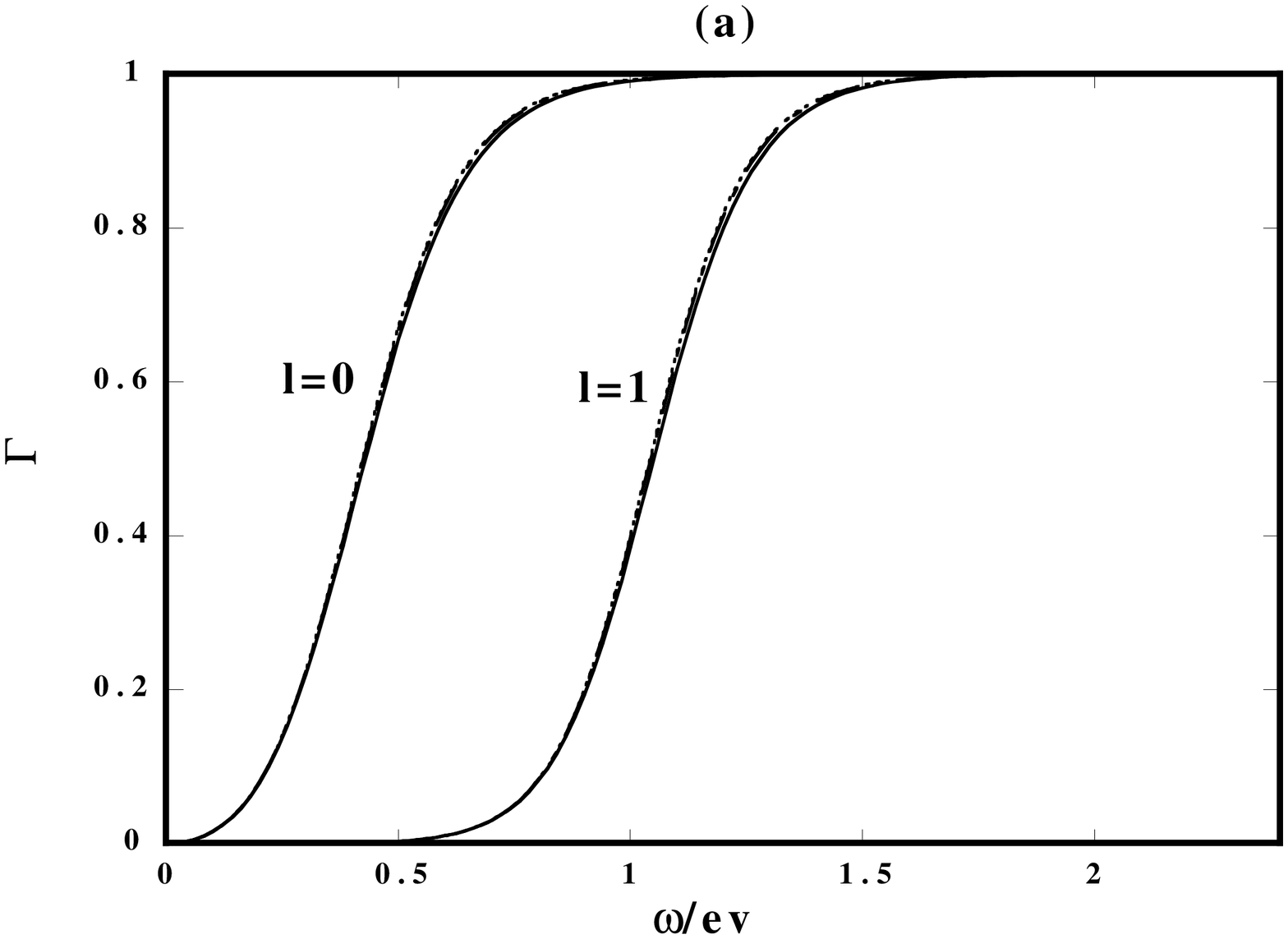}{} \\
\segmentfig{8cm}{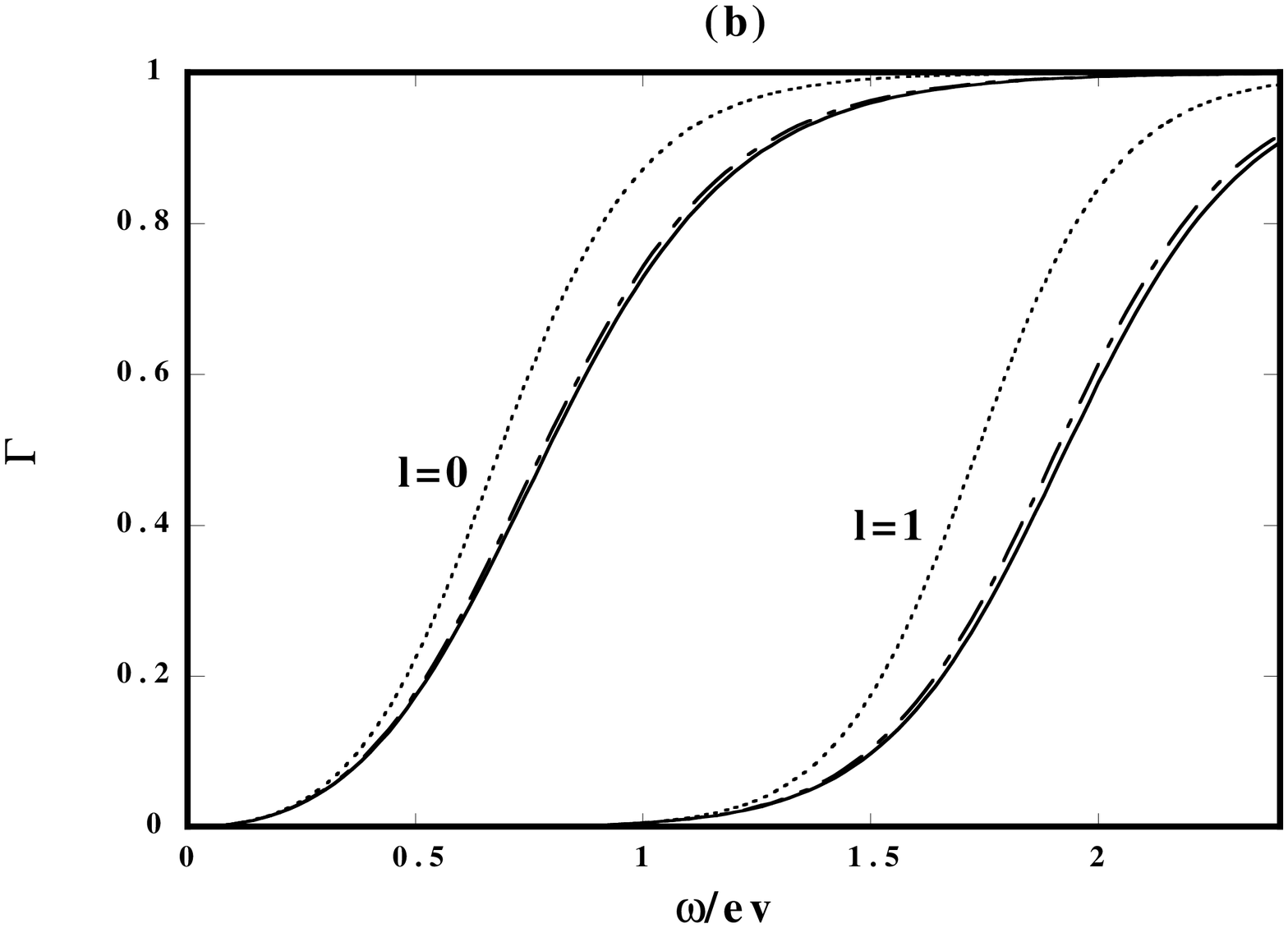}{}
\caption{Energy dependence of the transmission amplitude $\Gamma$ for 
monopole black holes with 
$\lambda /e^{2}=0.1$ (solid lines) and $\lambda /e^{2}=1$ 
(dot-dashed lines) and  of RN black holes (dotted lines) for
$l=0$, $1$ mode which make the main contributions to the  Hawking
radiation. We choose $v/M_{pl}=0.05$ in these diagrams and  (a)
$r_{H}/(ev)^{-1}=0.55$ (b) $r_{H}/(ev)^{-1}=0.3$. Though the RN
black hole has the largest $\Gamma$ among  them, these are almost
indistinguishable in (a).  But when horizon radius becomes
small,  their difference becomes large because YM field and Higgs
field outside the horizon becomes large and  contribute to the
black hole structure. 
\label{transmission}
}
\end{figure}
\begin{figure}[htbp]
\segmentfig{8cm}{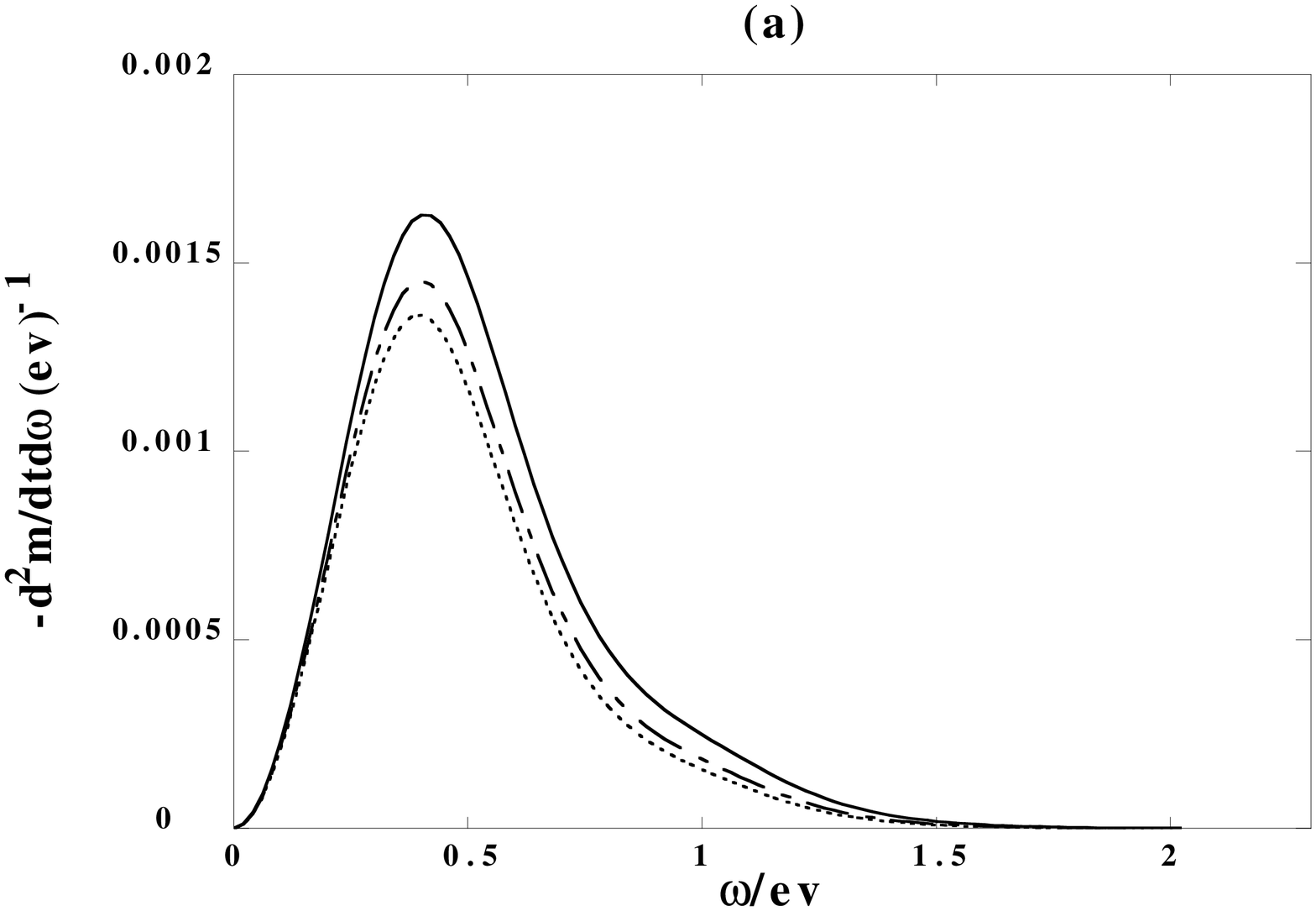}{} \\
\segmentfig{8cm}{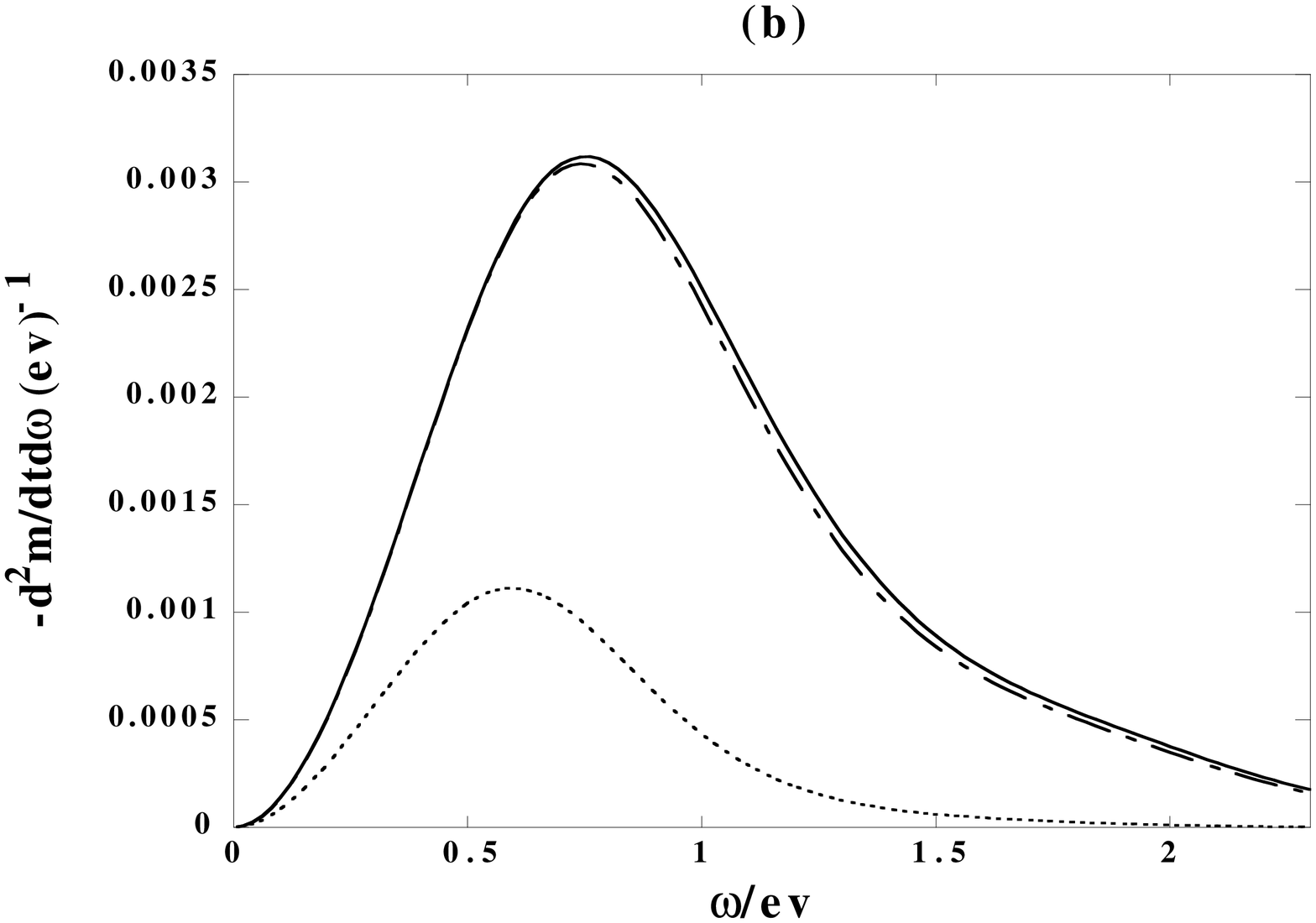}{}
\caption{Energy spectrum of the $-d^{2}m/dtd\omega$ for 
black holes corresponding to Fig. 3 (a), (b). 
They shows that why we can neglect the contribution from $l\geq 2$. 
\label{spectrum}
}
\end{figure}
\begin{figure}
\begin{center}
\segmentfig{8cm}{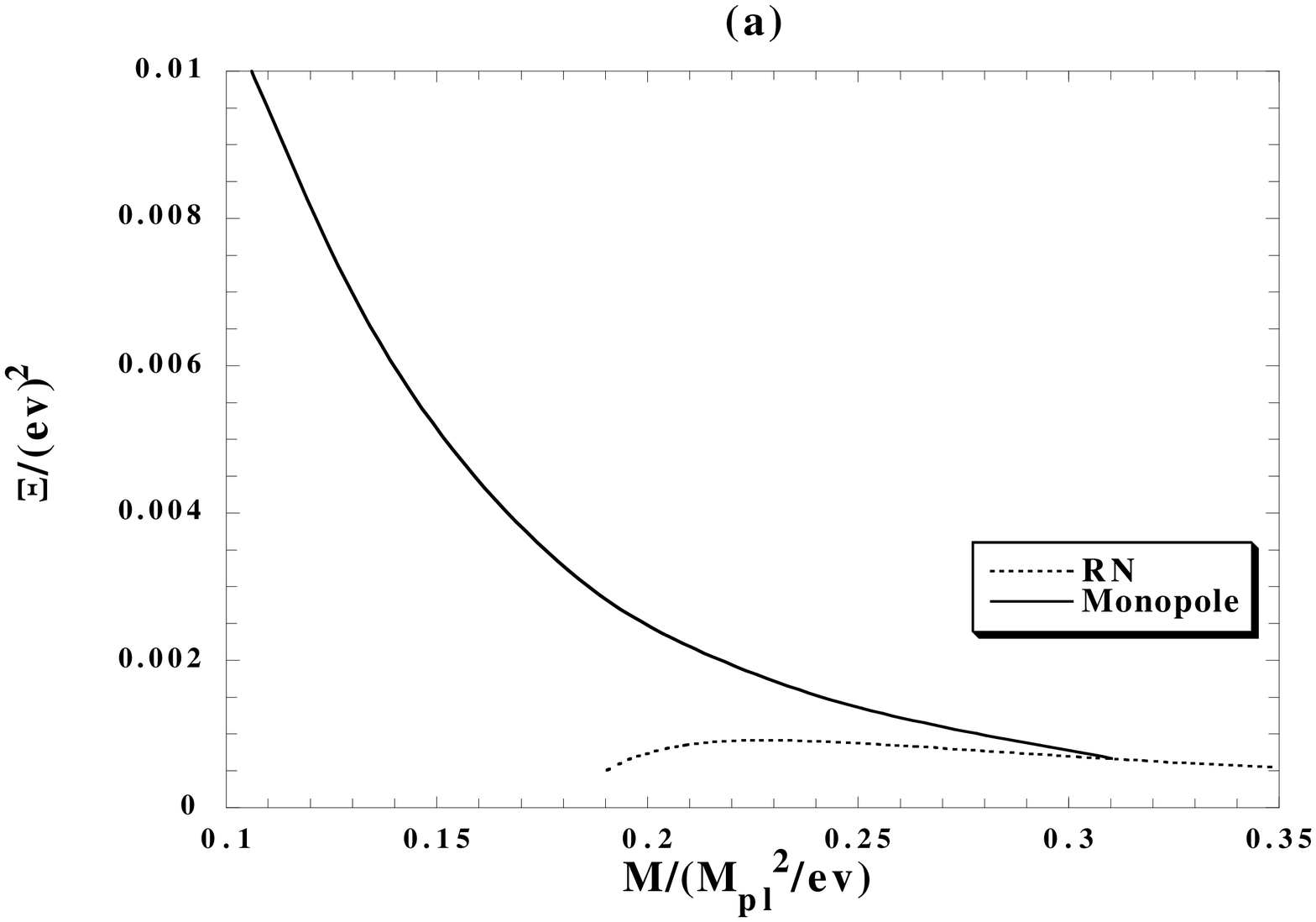}{}  \\
\segmentfig{8cm}{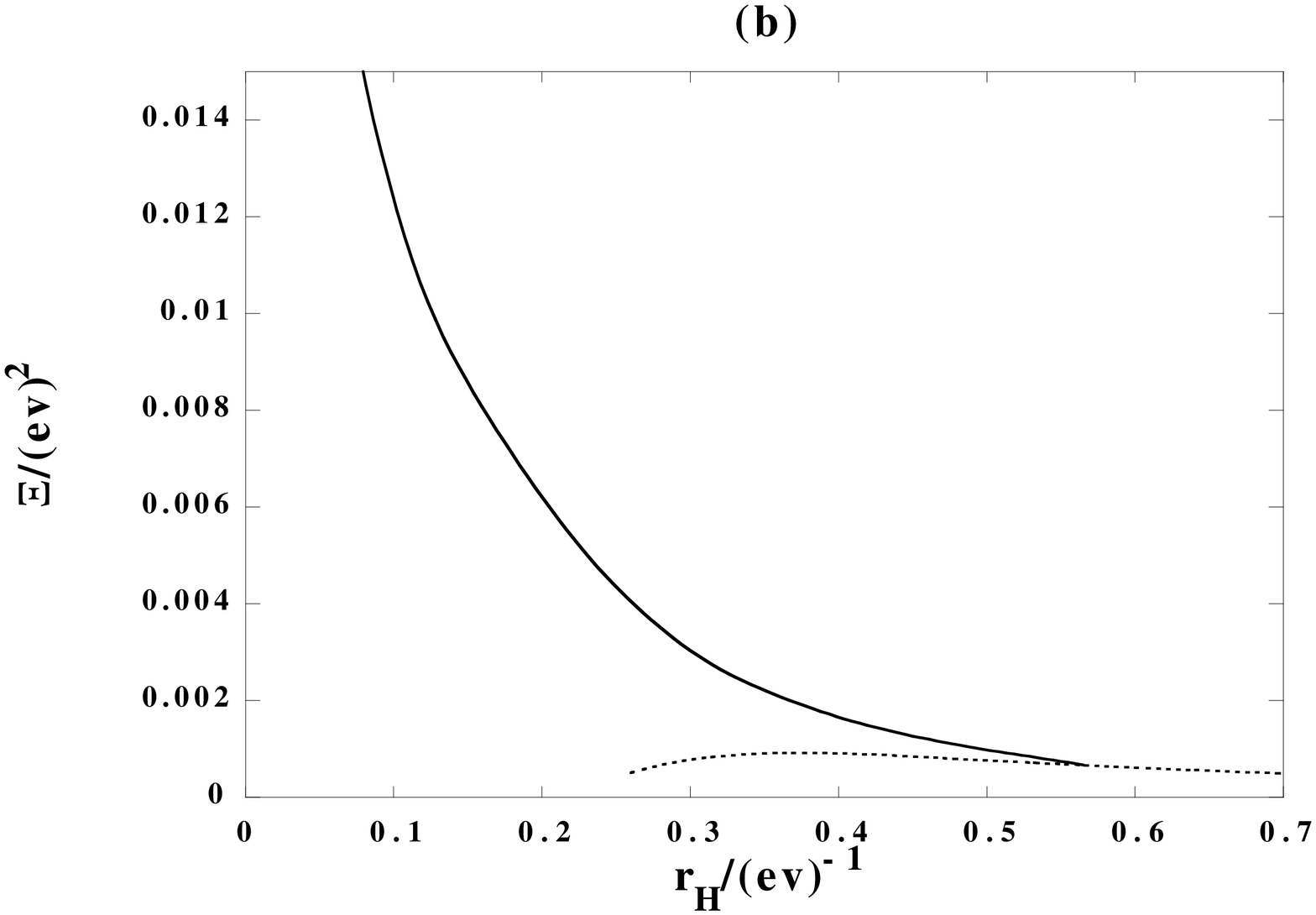}{}
\caption{The emission rate $\Xi /ev$ (a) in terms of the gravitational mass 
$M/(M_{pl}^{2}/ev)$ (b) in terms of the horizon radius $r_{H}/(ev)^{-1}$ 
for RN and monopole black holes for $v/M_{pl}=0.05$, 
$\lambda /e^{2}=1$.  If we assume the charge is conserved,  
though the RN black hole will stop  evaporating  at the extreme
limit, monopole black hole does not stop evaporating as in  the
Schwarzschild black hole. If we consider the effects of
quantum gravity, the results near $r_{H}\sim l_{p}$ may be changed. 
\label{evapo1}
}
\end{center}
\end{figure}
\begin{figure}
\begin{center}
\singlefig{8cm}{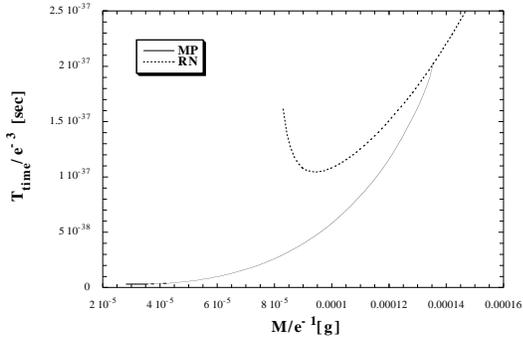}
\caption{The evaporation time scale $T_{time}$ in terms of the 
gravitational mass $M$ in the CGS units for the same solutions in Fig. 5. 
\label{timescale} 
}
\end{center}
\end{figure}
\begin{figure}
\begin{center}
\singlefig{8cm}{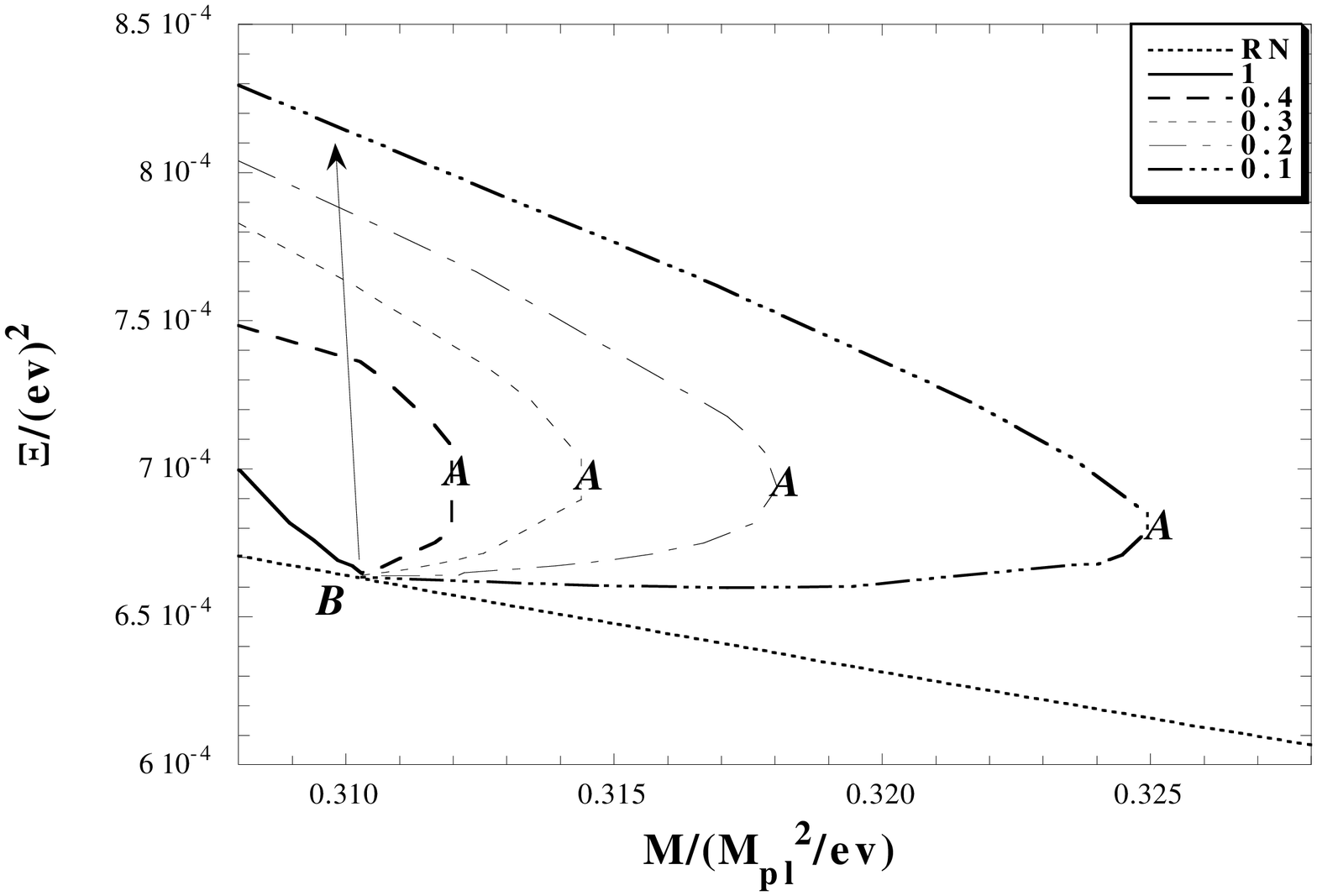}
\caption{The emission rate $\Xi /ev$ in terms of the gravitational  mass
$M/(M_{pl}^{2}/ev)$ for   RN and monopole black holes for
$v/M_{pl}=0.05$, 
$\lambda /e^{2}=0.1$, $0.2$, $0.3$, $0.4$, $1$ near the 
bifurcation point $B$.  The lines from $B$ to $A$ correspond 
to the emission rate of the monopole black hole which is thought 
to be unstable. So when the transition from  the RN black hole to
the monopole black hole occurs, emission rate will raise above
the points $A$.  
\label{evapo2}
}
\end{center}
\end{figure}
\begin{figure}
\begin{center}
\singlefig{8cm}{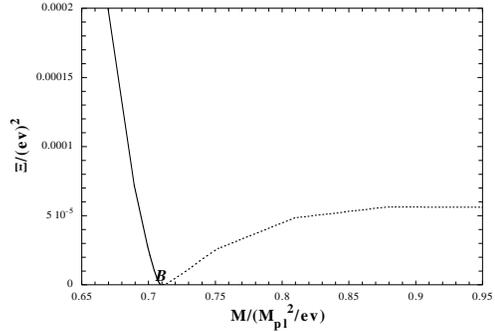}
\caption{The emission rate $\Xi /ev$ in terms of the gravitational mass 
$M/(M_{pl}^{2}/ev)$ for  
$v/M_{pl}=0.2$. We chose $\lambda /e^{2}=0.1$.  In this figure, 
we considered the situation bifurcation  point almost coincide
with extreme RN black  hole. This figure shows that evaporation
feature drastically changes near the extreme point because  of
the transition into the monopole black hole. 
\label{evapo3}
}
\end{center}
\end{figure}


\begin{thebibliography}{99}
\bibitem{footnote1}
We use the same notation as in \protect\cite{JK}.
\end{thebibliography}
\end{document}